# Packing, Phase Separation and Interface Compatibility in Reversibly Crosslinked Polymers

Abhishek S. Chankapure[1], Rahul Karmakar,[1] Srikanth Sastry[2], Sanat K. Kumar,[3] and Tarak K. Patra[1]

[1]Department of Chemical Engineering, Indian Institute of Technology Madras, Chennai, TN 600036, India
[2]Theoretical Sciences Unit and School of Advanced Materials, Jawaharlal Nehru Centre for Advanced Scientific Research, Bengaluru, KA 560064, India
[3]Department of Chemical Engineering, Columbia University, New York, NY 10027, USA

**Abstract:**

Vitrimers are a class of crosslinked polymer that are capable of undergoing bond exchange reactions, allowing structural reorganization while maintaining overall network integrity. Two key features that are particularly relevant when this vitrimer concept is used to compatibilize immiscible polymer blends are how they affect the (i) bulk polymer density and (ii) interfacial activity of the crosslink groups. To probe these issues, we model both a bulk polymer melt and a thin film of a polymer melt both with explicit small molecular crosslinkers, in the associative limit, i.e., when the number of crosslinks are fixed. We show that the bulk density and the distribution of stickers within a polymer matrix is strongly influenced by their size and interactions with the base polymer. Specifically, when the crosslinkers are chemically compatible with the base polymer, then the overall packing fraction increases, regardless of crosslinker size, while it decreases when crosslinkers are incompatible with the polymers. Similarly, the crosslinkers segregate preferentially to the polymer-air interface when they are incompatible with the polymer chains, leading to a reduction interfacial tension. Thus, these incompatible crosslinkers should help in affecting both the miscibility of polymer blends, and also their compatibility by creating copolymer structures at the interface. These results demonstrate the key role of crosslinker-polymer interactions and crosslinker size on the structural and interfacial properties of vitrimer melts.

Authors to correspond: TKP, E-mail: tpatra@iitm.ac.in
SKK, E-mail: sk2794@columbia.edu

# I. Introduction

Permanently crosslinked polymers are widely utilized for their excellent mechanical properties. However, their inability to be reprocessed makes them inappropriate in a circular polymer economy. In contrast, reversibly crosslinked polymers offer reprocessability and thus can circumvent these difficulties.[1–11] Dynamic bond exchange, which is driven by reversible chemical reactions such as disulfide exchange or transesterification, imparts advantageous properties including enhanced mechanical performance, tunable rheology, self-healing capability, and shape memory behavior.[12–16] Along these lines, recent experiments report that the surface tension[17] and mass density[18] of a polymer melt can be significantly impacted by dynamic crosslinks. Further, our recent findings demonstrated that reversible crosslinking can significantly influence the miscibility of polymer blends, potentially altering their phase behavior and compatibility.[19] The underpinning mechanisms by which these property changes occur on dynamically crosslinking a polymer melt remain poorly understood. Here, we ask how the size and interaction of added crosslinkers affect their distribution in a polymer matrix, and affect the bulk density and interfacial properties of the polymer.

Our previous simulations showed that introducing crosslinks increases the bulk density of a polymer matrix due to the loss of translational entropy.[20] This densification leads to a sharper gas-liquid interface and increased surface tension. In these simulations, the crosslinkers were modeled as phantom entities with no size, which is most applicable to systems where oxidative, thermal, or radiative crosslinking occurs in a polymer melt (i.e., without added crosslinking agents or reactive molecule additives which have their own volumes). Such densification and the resultant increase in surface tension does not offer a mechanism for the experimentally inferred reduction of surface tension[19], and suggests exploration of other mechanisms that may be at play. Here, we posit that, in case of explicit molecular-sized crosslinks, crosslink moiety-polymer interactions play dominant roles in determining the structure and properties of this class of materials, while steric effects play a secondary (but not unimportant) role. Along this line, in recent experiments, crosslink groups that are incompatible with base polymer matrixes are found to microphase separate and form percolated network structures in polymer matrixes.[21–23] We note that prior computer simulations of molecular crosslinks primarily focused on their role in the dynamics and rheological properties of polymers when they are fully compatible with the base polymer.[24–31] Here, our objectives are to study the thermodynamically equilibrated bulk and interfacial properties of these materials with an emphasis on the compatability of crosslinkers with the polymer matrix, and the size of the crosslinker.

A polymer matrix containing molecular crosslinkers can, in the absence of crosslinking reactions, be conceptually mapped onto a nanoparticle–polymer composite, where the crosslinkers act as inert filler particles dispersed within the polymer matrix.[32,33] In this limiting case, the crosslinkers contribute primarily through excluded volume and non-bonded interactions (rather than through the formation of covalent or dynamic bonds with polymer chains). Based on this analogy, we hypothesize that the equilibrium thermodynamic and structural properties of molecularly crosslinked polymer systems will exhibit strong parallels to those of uncrosslinked nanoparticle–polymer composites, which have been studied extensively over the last few decades.[34] Specifically, features such as chain conformations, segmental packing, and local heterogeneity are expected to show analogous trends in the two systems, with the principal distinction arising only when crosslinking reactions are activated, thereby introducing network connectivity, and these topological constraints are absent in the nanoparticle–polymer composite. Hence we examine and establish these connections between these two systems - explicitly crosslinked polymer network and nanoparticle-polymer composite with the aim of improving the fundamental understanding of interface compatibility via inert and reactive nanoparticles.

The polymer is modeled as a coarse-grained chain of catenated monomers and crosslinkers are modeled separately as spherical particles. A crosslinker particle is covalently bonded to two different monomers of the host polymer matrix, Figure 1a. Therefore, the number of crosslink bonds is constant in the system. We consider two canonical cases - 1) the polymer-crosslink interaction is favorable, modeled by having all pairs interact through the same Lennard-Jones (LJ) potential (monomer-monomer, crosslinker-crosslinker, monomer-crosslinker) and 2) the polymer-crosslink interaction alone is unfavorable, i.e., the crosslinker-monomer interact through the Weeks-Chandler-Andersen (WCA) potential. We consider the associative bond exchange mechanism wherein a crosslink bond is swapped between two monomers. This crosslink bond exchange is simulated using a Monte Carlo (MC) scheme, while molecular dynamics (MD) simulations are conducted to relax the system. We also perform MD simulations of polymer-particle mixture without crosslink bonds. We find that the bulk packing fraction of the composite increases monotonically with the fraction of crosslinks and their size when polymer-crosslink interactions are favorable. The bulk packing fraction decreases only when polymer-crosslink interactions are unfavorable. We also find that the crosslinkers tend to aggregate preferentially at the interface when crosslink-polymer interaction is unfavorable, thereby decreasing the surface tension. These crosslink bonds are found to shift the equilibrium density of the particle-polymer blend towards higher packing fraction. Our findings strongly suggest that the incorporation of bulkier or chemically dissimilar crosslinkers likely enhances their ability to compatibilize immiscible blends.

## II. Model and Methodology

We employ the Kremer-Grest bead-spring polymer model[35] and perform hybrid MD – MC simulations. Two non-bonded monomers interact via the Lennard-Jones (LJ) potential of the form $V(r) = 4\epsilon \left[\left(\frac{\sigma}{r}\right)^{12} - \left(\frac{\sigma}{r}\right)^{6}\right]$, truncated and shifted to zero at a cutoff distance $r_c = 2.5\sigma$. Here, $\sigma$ is the monomer diameter, while the cohesive interaction strength between two monomers is $\epsilon$. Catenated monomers on a chain are held together by the standard finitely extensible nonlinear elastic (FENE) potential $V_{bond}(r_{ij}) = -0.5kR_0^2 \ln\left[1 - \left(\frac{r_{ij}}{R_0}\right)^2\right] + 4\epsilon \left[\left(\frac{\sigma}{r}\right)^{12} - \left(\frac{\sigma}{r}\right)^{6}\right] + \epsilon$. The second term in $V_{bond}(r_{ij})$ alone is truncated at $r = 2^{1/6}\sigma$. The $R_0 = 1.5\sigma$, and $k = 30\epsilon$. The number of monomers in a chain is 20. The crosslinker particles are also modelled as spherical beads, with their sizes being varied from $d = 0.5\sigma$ to $3.0\sigma$ in a series of simulations. In separate simulations we treat crosslinkers which are energetically athermal with the chain monomers or they are incompatible. Crosslinker-crosslinker interactions were modeled through the LJ potential. In the athermal case, we model the interaction between a monomer and a crosslinker via the LJ interaction with a cut-off distance of $2.5\sigma_{ij}$, where $\sigma_{ij} = (\sigma_i + \sigma_j)/2$ is the average size of the two particles. In the case of an incompatible crosslinker, we model the crosslinker-monomer interaction via the WCA potential, which is the LJ interaction with a cutoff distance of $2^{1/6}\sigma_{ij}$. The bonds formed by the crosslinker with two separate monomers in the system, Figure 1a, are individually modeled via the FENE potential. While $R_0 = 1.5\sigma_{ij}$ [where $\sigma_{ij} = (d + \sigma)/2$ ], the bond constant is chosen such that $k\sigma_{ij}^2 = $ constant=30 for all the case studied. This ensures that the average equilibrium bond length between two bonded particles is $0.96\sigma_{ij}$ (cf. Figure 1b and c). [On the other hand, simply assigning $k$=30 for all crosslinker sizes causes the bond length normalized by $\sigma_{ij}$ to vary with changes in $d$, Figure 1b.] For comparison with our previous work we also use the phantom (or implicit) crosslinker model, where two monomers are directly connected by a FENE bond with $k\sigma_{ij}^2 = 30$ and $R_0 = 1.5\sigma_{ij}$. The total number of monomers in the simulation box is 10000. The number of crosslinkers is varied from 20 to 100. The simulation box is periodic in all three directions. The bulk system simulations are performed in the isothermal-isobaric ensemble, while the free standing film simulations are done in a canonical ensemble. For the free standing film simulations, one box dimension (z-axis) is significantly longer than the other two ( x- and y- axes), so that we model a thin film surrounded by its vapor along the z direction. We model an associative polymer network as schematically shown in Figure 1d, where the number of crosslink bonds is fixed and they are exchanged between monomers at an elevated temperature. The crosslinkers can attach to any monomer in the system, with the proviso that no monomer can have more than a single crosslink. In our simulations, the exchange of crosslink bonds between monomers are performed using a configurational-biased MC algorithm as implemented in our previous studies.[36] An MC move is performed to replace one of the monomers of a randomly selected crosslink bond by a neighboring uncrosslinked monomer in each attempt. We perform 200 MC moves (defined as an MC cycle) at regular intervals during the MD simulation conducted using the LAMMPS open-source code.[37] The MD time between two consecutive MC cycles is chosen to be $5\tau$. Here, $\tau = \sqrt{m\sigma^2/\epsilon}$ is the unit of time and $m$ is the monomer mass. The monomers and crosslinkers only move during the MD, and are not allowed to move during bond swap steps. The initial configurations are prepared by randomly inserting polymer chains and particles in a simulation box, and forming crosslink bonds between the filler particles and nearby monomers. Each filler particle is connected to two monomers via the crosslink bonds. We first perform $10^7$ MD steps with an integration timestep of $0.005\tau$ to equilibrate the system, followed by a production cycle of $10^7$ MD steps. The duration of our simulations is long enough to relax the chains as shown in Figure 1e for few representative cases. The reduced temperature of the system $\left(\frac{k_BT}{\epsilon}\right) = 1$ is maintained by the Nosé-Hover thermostat for all the simulations. For bulk system simulations, the reduced pressure $\left(\frac{P\sigma^3}{\epsilon}\right) = 1$ is maintained by the Nosé-Hoover barostat, while the film simulations naturally evolve to their vapor pressures at the temperature of interest (which turns out to be close to zero pressure). Here, $T$, $P$ and $k_B$ are the temperature, pressure and Boltzmann constant, respectively.

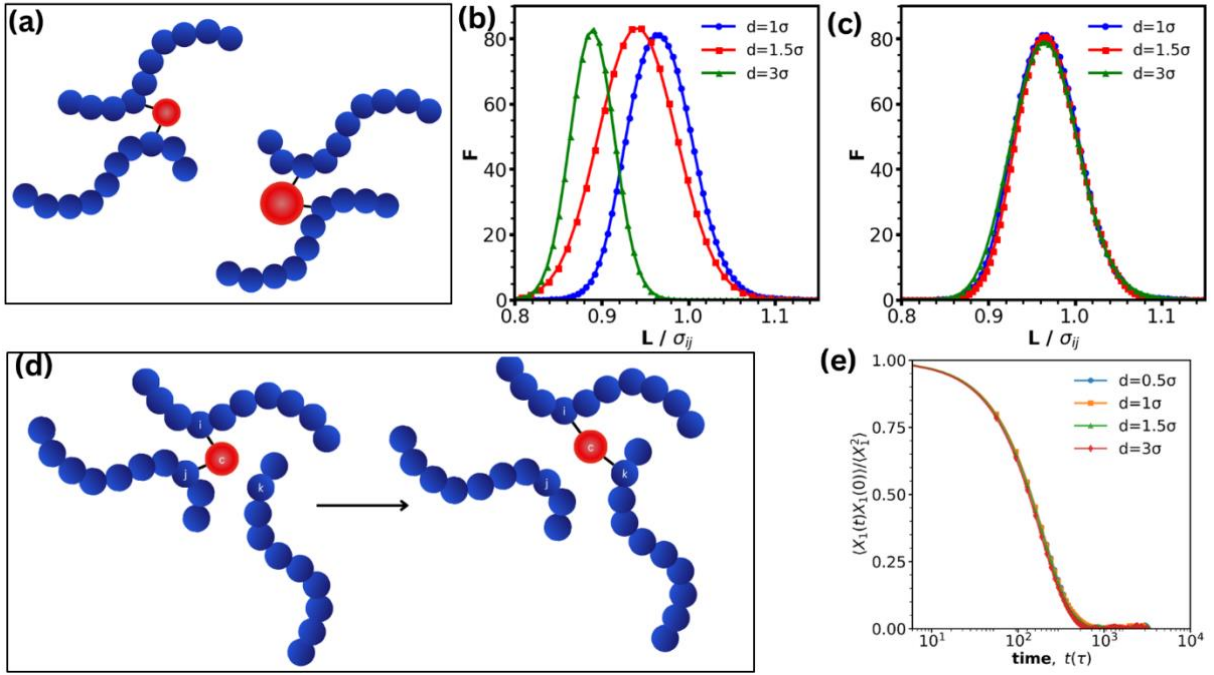

*Figure 1: (a) Schematic representations of crosslinked polymers with two different sizes of the crosslinker (blue bead). Connected read beads are monomers of a polymer chain. (b) Crosslink bond length distribution for different crosslinker sizes when $R_0 = 1.5\sigma_{ij}$ and $k = 30\epsilon$. (c) Crosslink bond length distribution when $k\sigma_{ij}^2 = 30 = constant$. (d) A schematic representation of dynamic bond exchange mechanism, where crosslinkers swap bonding partners in the network. (e) Longest Rouse mode autocorrelation function is shown as a function of time for different crosslinker size.*

## III. Results and Discussion

We first study the distribution of crosslinkers in the bulk phase when they are compatible with the polymer matrix. The pair correlation function for crosslinkers, $g_{c-c}(r)$ are shown in Figure 2(a-c) for three crosslink percentages - 2%, 10% and 20%, respectively. The crosslink percentage is defined as the proportion of monomers of the system that are connected via crosslink particles. For each of these crosslink concentrations, we also simulate the implicit model where two monomers are connected by FENE bonds without any explicit crosslinker particle. For the implicit case, we always observe good spatial dispersion of crosslink bonds, consistent with our previous study.[20] Moving from implicit to explicit crosslinkers, we observe that the $g_{c-c}(r)$ peak height increases as the crosslinker size increases. We also provide MD snapshots of the system for three crosslink concentrations, i.e., 2%, 10%, 20%, in Figure 2d-f. While implicit crosslinkers show gas-like pair correlation in the polymer matrix, explicit crosslinkers exhibit liquid-like pair distributions, particularly at higher concentrations and larger crosslinker sizes.

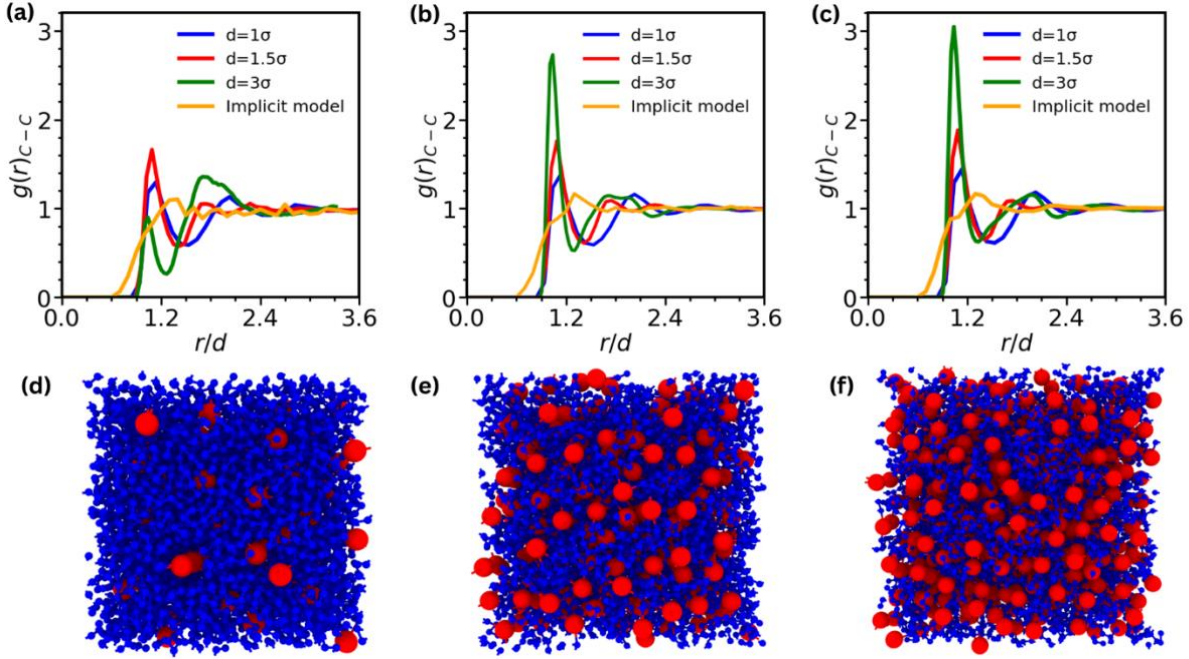

*Figure 2: The crosslink pair correlation function $g_{c-c}(r)$ for crosslinking concentration 2%, 10%, and 20% are shown in (a), (b), and (c), respectively, for varying crosslinker sizes along with the implicit model. MD snapshots of the system with 10% crosslinks are shown for 20%, 10%, and 20% in (d), (e) and (f), respectively. Red beads are crosslinks and blue beads are polymer beads.*

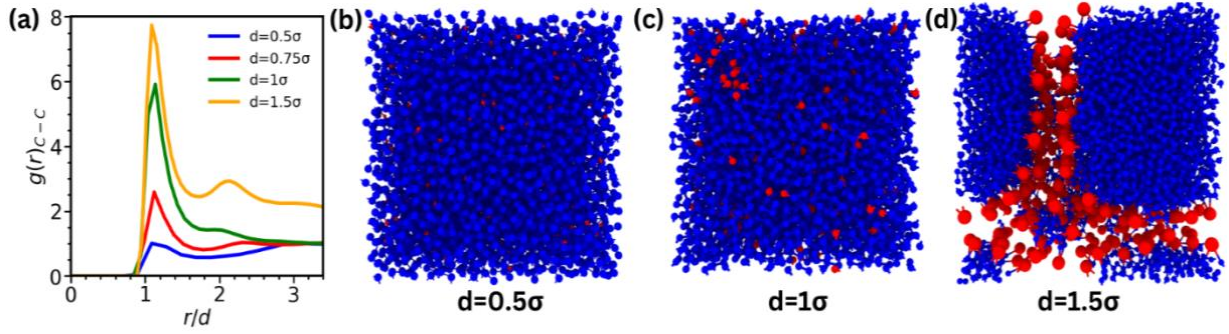

*Figure 3: (a) Crosslink pair correlation function $g_{c-c}(r)$ is shown for different WCA crosslinker size. The MD snapshots of the system for d=0.5σ, 0.75σ, and 1.5σ, are shown in (b), (c), and (d), respectively. Red beads are monomers and blue beads are crosslink particles.*

We now examine the distribution of incompatible crosslinkers, Figure 3a, for different crosslinker sizes at 10% concentration. MD snapshots for three such incompatible systems are reported in Figure 3b-d. We observe that for $d = 0.5\sigma$, there is hardly a first peak in the g(r), indicating uniform spatial dispersion of crosslinkers, in spite of the crosslinker-monomer interactions being unfavourable. This is very similar to the implicit crosslink distribution in the polymer matrix, Figure 2(a-c). The system progressively displays incompatibility as $d$ increases, leading to macrophase separation for $d_c \geq 1.5\sigma$. We infer that the extent of phase separation and the formation of hierarchical nanostructures in a dynamically crosslinked polymer network is governed by the bulkiness of the chemically dissimilar crosslink groups. In our present model, every monomer in a chain is eligible to participate in crosslink exchange reactions. If bond exchange is limited to specific "sticker" monomers, the extent of phase separation should change; under such an architecture, one would expect microphase separation of crosslinkers instead of a macrophase separation of all the crosslinkers shown in Figure 3d.[38]

Next, we calculate the packing fraction, which is defined as $\eta = \frac{\pi(N\sigma^3 + N_c d^3)}{6V}$, where $N$, $N_c$, and $d$ represent the number of monomer beads, number of crosslinkers (filler particles) and the size of a crosslinker, respectively. $V$ is the average system volume. $\eta$ quantifies how efficiently particles occupy space within the material. We first compute the packing fraction before the added particles crosslink the chain monomers. Under such conditions, the packing fraction ($\eta_0$) does not change significantly for smaller size filler particles as shown in Figure 4a. As the size of the filler particles increases, the packing fraction increases monotonically. Once we allow particles to crosslinks with the monomers and exchange their bonding partners, the packing fraction ($\eta$) system is increased in all cases, as shown in Figure 4a for 10% crosslinks. Moreover, the ratio of the two packing fractions, with crosslinks to the one without crosslinks varies linearly with the particle size and reaches a plateau for bigger size particle as shown in Figure 3b. Note that the y-axis varies from 1.01 to 1.04 as the filler particle size increases from $0.5\sigma$ to $3\sigma$ so that these changes are minimal. With this understanding, we plot the corresponding data for all of our systems in Figure 3c. For crosslinkers which are compatible with the polymer matrix, we observe a monotonic increase in $\eta$ as $d$ increases, irrespective of crosslinking density, with this monotonic increase only being reflective of the underlying base case of the nanocomposite. This suggests that larger LJ crosslinkers facilitate more efficient packing, possibly due to better space filling and reduced free volume. In contrast, crosslinkers which are incompatible with the base polymer matrix, result in a lower η which decreases monotonically with $d$. This reduction in packing efficiency is consistent with the phase separation that becomes increasingly prominent at larger crosslinker sizes. Since implicit crosslinks always increase the packing fraction, one can tune the packing fraction by choosing the size and compatibility of the crosslinkers with the base polymer matrix.

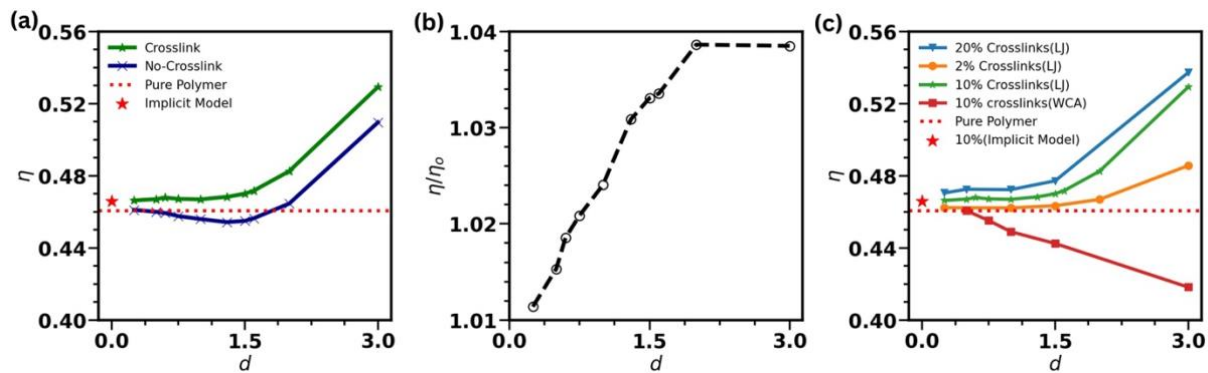

*Figure 4. Packing fraction (η) of the system for crosslinked and no-crosslink cases along with pure polymer and implicit crosslink model are shown in (a). The ratio of packing fractions of crosslink and no-crosslink cases are plotted as a function of the size of the crosslinker particle in (b). The (a, b) correspond to 10 % crosslinker particles. The packing fraction is plotted as a function of crosslinker size for various crosslink fractions including implicit model in (c)..*

To understand the consequence of changes in η on polymer properties, we investigate the role of explicit crosslinkers in tuning the surface properties of polymer thin films. As demonstrated in our previous work, the surface tension of the melt increases with implicit crosslinker fraction.[20] We reproduce this result in the present study and develop a comprehensive understanding by incorporating explicit crosslinkers with varying size and interactions. In the implicit model, crosslinkers are depleted from the interface and preferentially localized within the bulk region of a polymer film. In contrast, in the explicit model—especially for WCA interactions—the crosslinkers exhibit a clear tendency to accumulate at the interface, particularly as their size increases (Figure 5). This tendency is also found for the LJ case, but the effects is less pronounced. Our inert particle-polymer simulations reveal that the particles themselves spontaneously migrate toward the interface. This observation is consistent with previous findings in polymer nanocomposites, where interfaces are known to promote the segregation of nanoparticles due to entropic effects.[39–42] Therefore, we infer that the size of the dynamic crosslinkers can be used to tune this segregation tendency. As a result, the overall packing fraction near the interface increases significantly in the presence of LJ crosslinkers (Figure 4b), while it is notably suppressed in systems with WCA crosslinkers (Figure 4d). These contrasting trends underscore the role of crosslinker–polymer interactions and particle size in controlling the spatial distribution and packing behavior within thin films. Regardless, from a practical standpoint these crosslinkers spontaneously place themselves at interfaces, a fact that is particularly relevant when these moieties are used to compatibilize immiscible polymer blends.

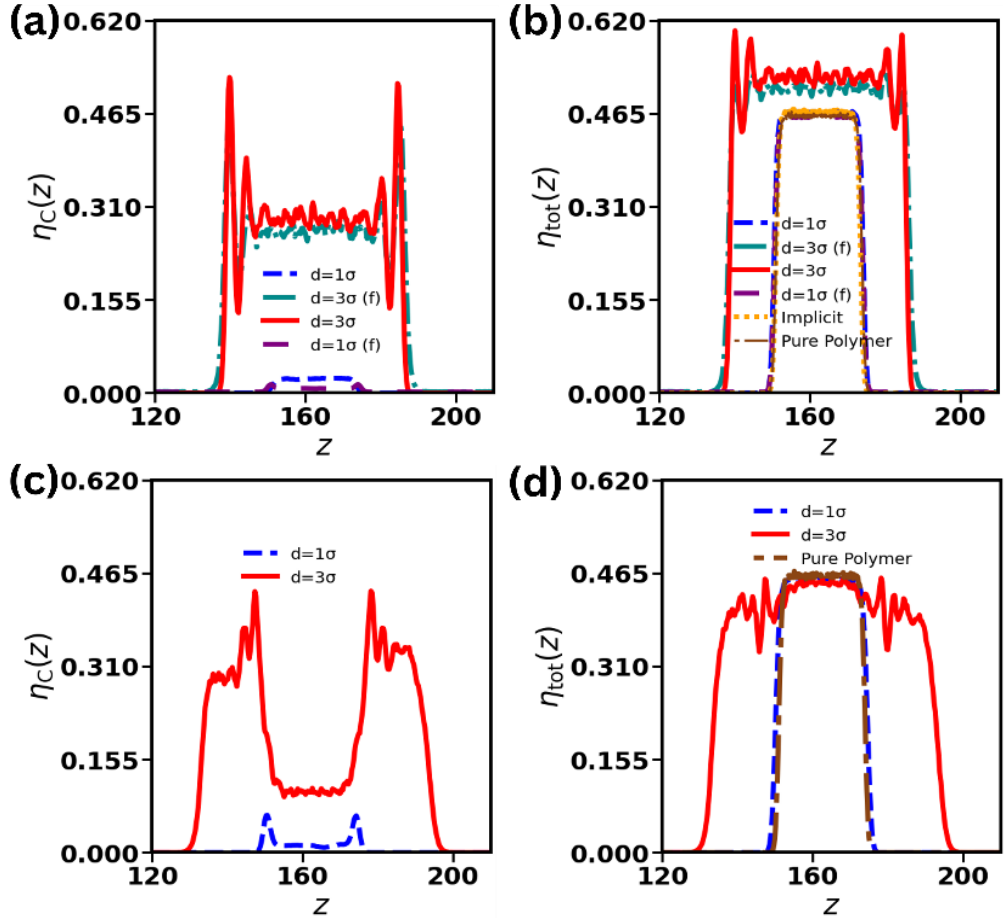

*Figure 5: Packing fraction of crosslinkers and all particles are plotted along the z-axis of the simulation box in (a) and (b), respectively, for the case of LJ crosslinkers. Similarly, crosslinker packing fraction and total packing fraction for the WCA case is shown in (c) and (d), respectively. All the data correspond to 10% crosslinks. The data sets, which are labelled as "f" correspond to free particles in a polymer film without any crosslink bonds between them and monomers.*

We estimate the interfacial tension, which is defined as $\gamma = \frac{L_z}{2}\left[p_{zz} - \frac{p_{xx}+p_{yy}}{2}\right]$, where $p_{xx}$, $p_{yy}$ and $p_{zz}$ are the pressure components along $x$, $y$, $z$ directions of the simulation box, respectively.[43] Here, $L_z$ is the box length along the $z$ direction. We calculate the components of the pressure tensor as $p_{ij} = \frac{1}{V}\sum_{k=1}^{M} m_k v_{ik} v_{jk} + \frac{1}{V}\sum_{k=1}^{M} r_{ik} f_{jk}$ where $v$, $r$ and $f$ denote the velocity, position and force components of a particle, respectively. The indices $i$ and $j$ refer to coordinate axes ($x$, $y$ and $z$), and the index $k$ runs over all the particles in the simulation box. The volume of the simulation box is $V$. We plot interfacial tension $\gamma$ (scaled with the value of pure polymer melt interfacial tension $\gamma_{pure}$) as a function of the size of a crosslinker in Figure 6. Here, $d_c = 0$ corresponds to the implicit case. The surface tension of the film increases for implicit crosslinks. Further, the surface tension increases for smaller size LJ crosslinkers ($d \cong 0.25$). As the size increases the surface tension starts to decrease. We also note that the change in surface tension for the case of free (inert) LJ particles, where they are not crosslinked with polymer chains, follows a similar trend as the crosslinked case. The WCA crosslinkers show the largest interfacial tension decrease. These results demonstrate that interfacial properties in dynamically crosslinked polymer networks is analogous to polymer-nanoparticle blends. It can be tuned by modulating both particle size and interaction with the base polymer matrix. The dynamic crosslinks serve as an additional handle to control the particle segregation and subsequently interfacial properties of polymers. From a design perspective, WCA crosslinkers and bulkier LJ crosslinkers are more efficient in compatibilizing the air-polymer interface relative to small LJ crosslinkers.

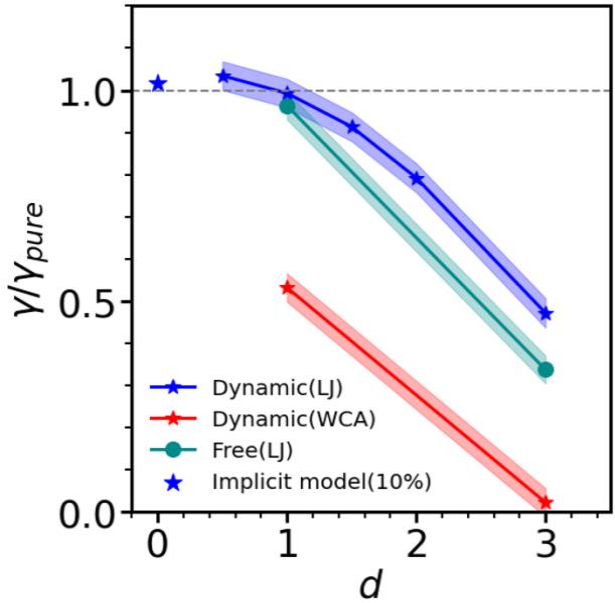

Figure 6: The normalised surface tension is plotted as a function of crosslinker size for LJ and WCA interaction along with implicit model for 10% crosslink concentrations. The data labeled as Free(LJ) represent the polymer-particle mixture without any crosslink bonds between them.

## IV. Conclusions

We systematically investigate how the size and chemical compatibility of molecular crosslinkers influence the structural and interfacial properties of reversibly crosslinked polymer networks using hybrid MC–MD simulations. When crosslinkers are chemically compatible with the polymer matrix (modeled via the LJ interaction), the bulk packing fraction increases monotonically with the crosslinker size and concentration – this result largely reflects the thermodynamics of the mixtures of polymers and nanoparticles without crosslinks. The crosslinks only slightly increase the packing fraction. In contrast, chemically incompatible crosslinkers (modeled via the WCA interaction) lead to a decrease in packing efficiency and eventually induce microphase and macrophase separation as their size increases. These effects are absent in the implicit crosslinking model, which represent crosslinkers as bonds between monomer pairs without volume or interaction. Further, larger and/or incompatible crosslinkers preferentially localize at polymer-air interfaces, resulting in a reduction of surface tension. This behavior contrasts with implicit crosslinking, which increases the surface tension due to the densification of the bulk region. Our results highlight that the interplay of crosslinker size and compatibility can be leveraged to finetune bulk density, phase behavior, and interfacial properties of polymers. This provides a rational design route for achieving desired functionalities such as tailored miscibility and interface compatibility. Overall, this study provides a systematic understanding of the role of crosslinking from implicit to bulker molecular crosslinks on the thermodynamically equilibrated properties of polymers, especially for applications involving heterogeneous structures and interfaces.

## Acknowledgements

This work is made possible by financial support from the SERB, DST, India through a core research grant (CRG/2022/006926).

## References

1. Montarnal, D., Capelot, M., Tournilhac, F. & Leibler, L. Silica-Like Malleable Materials from Permanent Organic Networks. *Science* **334**, 965–968 (2011).

2. Lu, Y.-X., Tournilhac, F., Leibler, L. & Guan, Z. Making Insoluble Polymer Networks Malleable via Olefin Metathesis. *J. Am. Chem. Soc.* **134**, 8424–8427 (2012).

3. Snyder, R. L., Fortman, D. J., De Hoe, G. X., Hillmyer, M. A. & Dichtel, W. R. Reprocessable Acid-Degradable Polycarbonate Vitrimers. *Macromolecules* **51**, 389–397 (2018).

4. Jin, Y., Yu, C., Denman, R. J. & Zhang, W. Recent advances in dynamic covalent chemistry. *Chem. Soc. Rev.* **42**, 6634–6654 (2013).

5. Samanta, S., Kim, S., Saito, T. & Sokolov, A. P. Polymers with Dynamic Bonds: Adaptive Functional Materials for a Sustainable Future. *J. Phys. Chem. B* **125**, 9389–9401 (2021).

6. Maaz, M., Riba-Bremerch, A., Guibert, C., Van Zee, N. J. & Nicolaÿ, R. Synthesis of Polyethylene Vitrimers in a Single Step: Consequences of Graft Structure, Reactive Extrusion Conditions, and Processing Aids. *Macromolecules* **54**, 2213–2225 (2021).

7. Long, R., Qi, H. J. & Dunn, M. L. Modeling the mechanics of covalently adaptable polymer networks with temperature-dependent bond exchange reactions. *Soft Matter* **9**, 4083–4096 (2013).

8. Meng, F., Pritchard, R. H. & Terentjev, E. M. Stress Relaxation, Dynamics, and Plasticity of Transient Polymer Networks. *Macromolecules* **49**, 2843–2852 (2016).

9. Ricarte, R. G. & Shanbhag, S. Unentangled Vitrimer Melts: Interplay between Chain Relaxation and Cross-link Exchange Controls Linear Rheology. *Macromolecules* **54**, 3304–3320 (2021).

10. Semenov, A. N. & Rubinstein, M. Thermoreversible Gelation in Solutions of Associative Polymers. 1. Statics. *Macromolecules* **31**, 1373–1385 (1998).

11. Rubinstein, M. & Semenov, A. N. Thermoreversible Gelation in Solutions of Associating Polymers. 2. Linear Dynamics. *Macromolecules* **31**, 1386–1397 (1998).

12. Porath, L., Soman, B., Jing, B. B. & Evans, C. M. Vitrimers: Using Dynamic Associative Bonds to Control Viscoelasticity, Assembly, and Functionality in Polymer Networks. *ACS Macro Lett.* 475–483 (2022) doi:10.1021/acsmacrolett.2c00038.

13. Zou, Z. *et al.* Rehealable, fully recyclable, and malleable electronic skin enabled by dynamic covalent thermoset nanocomposite. *Sci. Adv.* **4**, eaaq0508 (2018).

14. Zhang, V., Kang, B., Accardo, J. V. & Kalow, J. A. Structure–Reactivity–Property Relationships in Covalent Adaptable Networks. *J. Am. Chem. Soc.* **144**, 22358–22377 (2022).

15. Zhang, Z., Chen, Q. & Colby, R. H. Dynamics of associative polymers. *Soft Matter* **14**, 2961–2977 (2018).

16. Jiang, N., Zhang, H., Yang, Y. & Tang, P. Molecular dynamics simulation of associative polymers: Understanding linear viscoelasticity from the sticky Rouse model. *J. Rheol.* **65**, 527–547 (2021).

17. Zhao, W., Zhou, J., Hu, H., Xu, C. & Xu, Q. The role of crosslinking density in surface stress and surface energy of soft solids. *Soft Matter* **18**, 507–513 (2022).

18. Lewis, B., Dennis, J. M., Park, C. & Shull, K. R. Glassy Dynamics of Epoxy-Amine Thermosets Containing Dynamic, Aromatic Disulfides. *Macromolecules* **57**, 7112–7122 (2024).

19. Clarke, R. W. *et al.* Dynamic crosslinking compatibilizes immiscible mixed plastics. *Nature* **616**, 731–739 (2023).

20. Karmakar, R. *et al.* Computer simulations of entropic cohesion in reversibly crosslinked polymers. *Soft Matter* **21**, 348–355 (2025).

21. Ricarte, R. G., Tournilhac, F. & Leibler, L. Phase Separation and Self-Assembly in Vitrimers: Hierarchical Morphology of Molten and Semicrystalline Polyethylene/Dioxaborolane Maleimide Systems. *Macromolecules* **52**, 432–443 (2019).

22. Arbe, A. *et al.* Microscopic Evidence for the Topological Transition in Model Vitrimers. *ACS Macro Lett.* **12**, 1595–1601 (2023).

23. Polgar, L. M. *et al.* Effect of Rubber Polarity on Cluster Formation in Rubbers Cross-Linked with Diels–Alder Chemistry. *Macromolecules* **50**, 8955–8964 (2017).

24. Zhao, H. *et al.* Unveiling the Multiscale Dynamics of Polymer Vitrimers Via Molecular Dynamics Simulations. *Macromolecules* **56**, 9336–9349 (2023).

25. Zhao, H. *et al.* Molecular Dynamics Simulation of the Structural, Mechanical, and Reprocessing Properties of Vitrimers Based on a Dynamic Covalent Polymer Network. *Macromolecules* (2022) doi:10.1021/acs.macromol.1c02034.

26. Perego, A. & Khabaz, F. Volumetric and Rheological Properties of Vitrimers: A Hybrid Molecular Dynamics and Monte Carlo Simulation Study. *Macromolecules* **53**, 8406–8416 (2020).

27. Nie, W., Douglas, J. F. & Xia, W. Competing Effects of Molecular Additives and Cross-Link Density on the Segmental Dynamics and Mechanical Properties of Cross-Linked Polymers. *ACS Eng. Au* **3**, 512–526 (2023).

28. Perego, A. & Khabaz, F. Effect of bond exchange rate on dynamics and mechanics of vitrimers. *J. Polym. Sci.* **59**, 2590–2602 (2021).

29. Xia, J., Kalow, J. A. & Olvera de la Cruz, M. Structure, Dynamics, and Rheology of Vitrimers. *Macromolecules* **56**, 8080–8093 (2023).

30. Xia, J. & de la Cruz, M. O. Effect of molecular structure on the dynamics and viscoelasticity of vitrimers. *Polymer* **308**, 127371 (2024).

31. Lin, T.-W., Mei, B., Dutta, S., Schweizer, K. S. & Sing, C. E. Molecular Dynamics Simulation and Theoretical Analysis of Structural Relaxation, Bond Exchange Dynamics, and Glass Transition in Vitrimers. *Macromolecules* **58**, 1481–1497 (2025).

32. Kumar, S. K. & Krishnamoorti, R. Nanocomposites: Structure, Phase Behavior, and Properties. *Annu. Rev. Chem. Biomol. Eng.* **1**, 37–58 (2010).

33. Jancar, J. *et al.* Current issues in research on structure–property relationships in polymer nanocomposites. *Polymer* **51**, 3321–3343 (2010).

34. Kumar, S. K., Benicewicz, B. C., Vaia, R. A. & Winey, K. I. 50th Anniversary Perspective: Are Polymer Nanocomposites Practical for Applications? *Macromolecules* **50**, 714–731 (2017).

35. Kremer, K. & Grest, G. S. Molecular dynamics (MD) simulations for polymers. *J. Phys. Condens. Matter* **2**, SA295–SA298 (1990).

36. Karmakar, R., Sastry, S., Kumar, S. & Patra, T. K. Rouse Mode Analysis of Chain Relaxation in Reversibly Crosslinked Polymer Melts. *Phys. Chem. Chem. Phys.* (2025) doi:10.1039/D5CP00632E.

37. Thompson, A. P. *et al.* LAMMPS - a flexible simulation tool for particle-based materials modeling at the atomic, meso, and continuum scales. *Comput. Phys. Commun.* **271**, 108171 (2022).

38. Karmakar, R., Chankapure, A. S., Sastry, S., Kumar, S. K. & Patra, T. K. Microphase Separation Controls the Dynamics of Associative Vitrimers. Preprint at https://doi.org/10.48550/arXiv.2506.21066 (2025).

39. Sriramoju, K. K. & Padmanabhan, V. Controlling the Location of Bare Nanoparticles in Polymer-Nanoparticle Blend Films by Adding Polymer-Grafted Nanoparticles. *Phys. Rev. Lett.* **114**, 258301 (2015).

40. McGarrity, E. S., Frischknecht, A. L., Frink, L. J. D. & Mackay, M. E. Surface-Induced First-Order Transition in Athermal Polymer-Nanoparticle Blends. *Phys. Rev. Lett.* **99**, 238302 (2007).

41. Frischknecht, A. L., Padmanabhan, V. & Mackay, M. E. Surface-induced phase behavior of polymer/nanoparticle blends with attractions. *J. Chem. Phys.* **136**, 164904 (2012).

42. Lee, J. Y., Shou, Z. & Balazs, A. C. Predicting the Morphologies of Confined Copolymer/Nanoparticle Mixtures. *Macromolecules* **36**, 7730–7739 (2003).

43. Meenakshisundaram, V., Hung, J.-H., Patra, T. K. & Simmons, D. S. Designing Sequence-Specific Copolymer Compatibilizers Using a Molecular-Dynamics-Simulation-Based Genetic Algorithm. *Macromolecules* **50**, 1155–1166 (2017).